\documentclass[letterpaper]{article} 
\usepackage{aaai24}  
\usepackage{times}  
\usepackage{helvet}  
\usepackage{courier}  
\usepackage[hyphens]{url}  
\usepackage{graphicx} 
\usepackage{multirow,booktabs,amssymb,amsmath}
\usepackage{natbib}  
\usepackage{caption} 
\usepackage{algorithm}
\usepackage{algorithmic}
\usepackage{color}
\usepackage{threeparttable}
\usepackage{newfloat}
\usepackage{listings}
\DeclareCaptionStyle{ruled}{labelfont=normalfont,labelsep=colon,strut=off} 
\floatstyle{ruled}
\newfloat{listing}{tb}{lst}{}
\floatname{listing}{Listing}
\title{A Multi-Modal Contrastive Diffusion Model for Therapeutic Peptide Generation}

\author {
    Yongkang Wang\textsuperscript{\rm 1}\equalcontrib ,
    Xuan Liu\textsuperscript{\rm 1}\equalcontrib ,
    Feng Huang\textsuperscript{\rm 1},
    Zhankun Xiong\textsuperscript{\rm 1},
    Wen Zhang\textsuperscript{\rm 1,\rm 2 \rm 3}\thanks{Corresponding authors.}
}
\affiliations {
    \textsuperscript{\rm 1}College of Informatics, Huazhong Agricultural University, Wuhan 430070, China\\
    \textsuperscript{\rm 2}Hubei Key Laboratory of Agricultural Bioinformatics, Huazhong Agricultural University, Wuhan 430070, China\\
    \textsuperscript{\rm 3}Engineering Research Center of Intelligent Technology for Agriculture, Ministry of Education, Wuhan 430070,China\\

    \{wyky481, lx666, fhuang233, xiongzk\}@webmail.hzau.edu.cn, zhangwen@mail.hzau.edu.cn
}

\usepackage{bibentry}

\begin{document}

\maketitle
\begin{abstract}
Therapeutic peptides represent a unique class of pharmaceutical agents crucial for the treatment of human diseases. Recently, deep generative models have exhibited remarkable potential for generating therapeutic peptides, but they only utilize sequence or structure information alone, which hinders the performance in generation. In this study, we propose a Multi-Modal Contrastive Diffusion model (MMCD), fusing both sequence and structure modalities in a diffusion framework to co-generate novel peptide sequences and structures. Specifically, MMCD constructs the sequence-modal and structure-modal diffusion models, respectively, and devises a multi-modal contrastive learning strategy with inter-contrastive and intra-contrastive in each diffusion timestep, aiming to capture the consistency between two modalities and boost model performance. The inter-contrastive aligns sequences and structures of peptides by maximizing the agreement of their embeddings, while the intra-contrastive differentiates therapeutic and non-therapeutic peptides by maximizing the disagreement of their sequence/structure embeddings simultaneously. The extensive experiments demonstrate that MMCD performs better than other state-of-the-art deep generative methods in generating therapeutic peptides across various metrics, including antimicrobial/anticancer score, diversity, and peptide-docking.
\end{abstract}

\section{Introduction}
Therapeutic peptides, such as antimicrobial and anticancer peptides, are a unique class of pharmaceutical agents that comprise short chains of amino acids, exhibiting significant potential in treating complex human diseases \cite{jakubczykCurrentTrends2020}. Traditionally, therapeutic peptides are discovered through a comprehensive screening of sequence spaces using phage/yeast display technologies \cite{muttenthaler2021trends} or computational tools trained for scoring desired properties \cite{leeWhatCan2017, leeMachineLearningenabled2018}. However, the combinatorial space of possible peptides is vast and only a small solution satisfies therapeutic requirements; thus, such screening methods based on brute force can be time-consuming and costly.

In recent years, deep generative models (DGMs) have demonstrated success in generating images \cite{liuDesignGuidelines2022}, texts \cite{iqbalSurveyText2022}, proteins \cite{WU202118}, and also gained popularity in peptides. DGMs explored a more expansive chemical space that affords the creation of structurally novel peptides, by training neural networks to approximate the underlying distribution of observed or known ones \cite{wanDeepGenerative2022}. For example, autoregression-based methods depicted peptide sequences as sentences composed of residue tokens, so that the problem can be solved by predicting residue arrangement via recurrent neural networks (RNN) \cite{mullerRecurrentNeural2018, capecchiMachineLearning2021}. Variational autoencoder (VAE)-based methods generated new peptide sequences by sampling from the latent space learned through an encoder-decoder architecture, with or without therapeutic properties as conditional constraints \cite{ghorbaniDeepAttention2022, szymczak2023discovering}. Generative adversarial network (GAN)-based methods trained the generator and discriminator using known data, which compete against each other to generate new peptides \cite{tucsGeneratingAmpicillinLevel2020, oortAMPGANV2Guided2021, lin2022novo}. Nowadays, diffusion models \cite{yangDiffusionModels2023} are prevalent in the generation of protein sequences and structures, owing to their superior capability in fitting distributions compared to prior techniques \cite{shiProteinSequence2023, wuProteinStructure2022}. Likewise, these advanced diffusion models can be extended to peptide generation and are expected to deliver favorable outcomes.

Despite the commendable progress of efforts above, they focused on generating either sequences (i.e., residue arrangements) or structures (i.e., spatial coordinates of backbone atoms), ignoring that models fusing information from both modalities may outperform their uni-modal counterparts \cite{huang2021makes}. However, how to effectively integrate the multi-modal information and capture their consistency in peptide generation is a major challenge. Additionally, compared with generation tasks for images, texts, and proteins that involve millions of labeled samples, public datasets for therapeutic peptides typically contain only thousands of sequence or structure profiles, induced by the high cost of \textit{in vitro} screening. This limited amount of available data may result in overfitting \cite{webster2019detecting}, which confines generated outcomes within a restricted distribution, consequently compromising the model's generalization ability. How to fully leverage existing peptide data, such as therapeutic and non-therapeutic peptides, to enhance the generation performance could be regarded as another challenge. 

To address these challenges, we propose a \textbf{M}ulti-\textbf{M}odal \textbf{C}ontrastive \textbf{D}iffusion model for therapeutic peptide generation, named \textbf{MMCD}. Specifically, we build a multi-modal framework that integrates sequence-modal and structure-modal diffusion models for co-generating residue arrangements and backbone coordinates of peptides. To ensure consistency between the two modalities during the generation process, we bring in an inter-modal contrastive learning (Inter-CL) strategy. Inter-CL aligns sequences and structures, by maximizing the agreement between their embeddings derived from the same peptides at each diffusion timestep. Meanwhile, to avoid the issue of inferior performance caused by limited therapeutic peptide data, we incorporate substantial known non-therapeutic peptides as data augmentations to devise an intra-modal CL (Intra-CL). Intra-CL differentiates therapeutic and non-therapeutic peptides by maximizing the disagreement of their sequence/structure embeddings at each diffusion timestep, driving the model to precisely fit the distribution of therapeutic peptides. Overall, the main contributions of this work are described as follows:
\begin{itemize}
\item We propose a multi-modal diffusion model that integrates both sequence and structure information to co-generate residue arrangements and backbone coordinates of therapeutic peptides, whereas previous works focused only on a single modality.
\item We design the inter-intra CL strategy at each diffusion timestep, which aims to maximize the agreement between sequence and structure embeddings for aligning multi-modal information, and maximize the disagreement between therapeutic and non-therapeutic peptides for boosting model generalization.
\item Extensive experiments conducted on peptide datasets demonstrate that MMCD surpasses the current state-of-the-art baselines in generating therapeutic peptides, particularly in terms of antimicrobial/anticancer score, diversity, and pathogen-docking.
\end{itemize}

\begin{figure*}[ht]
\centering
\includegraphics[scale=0.74]{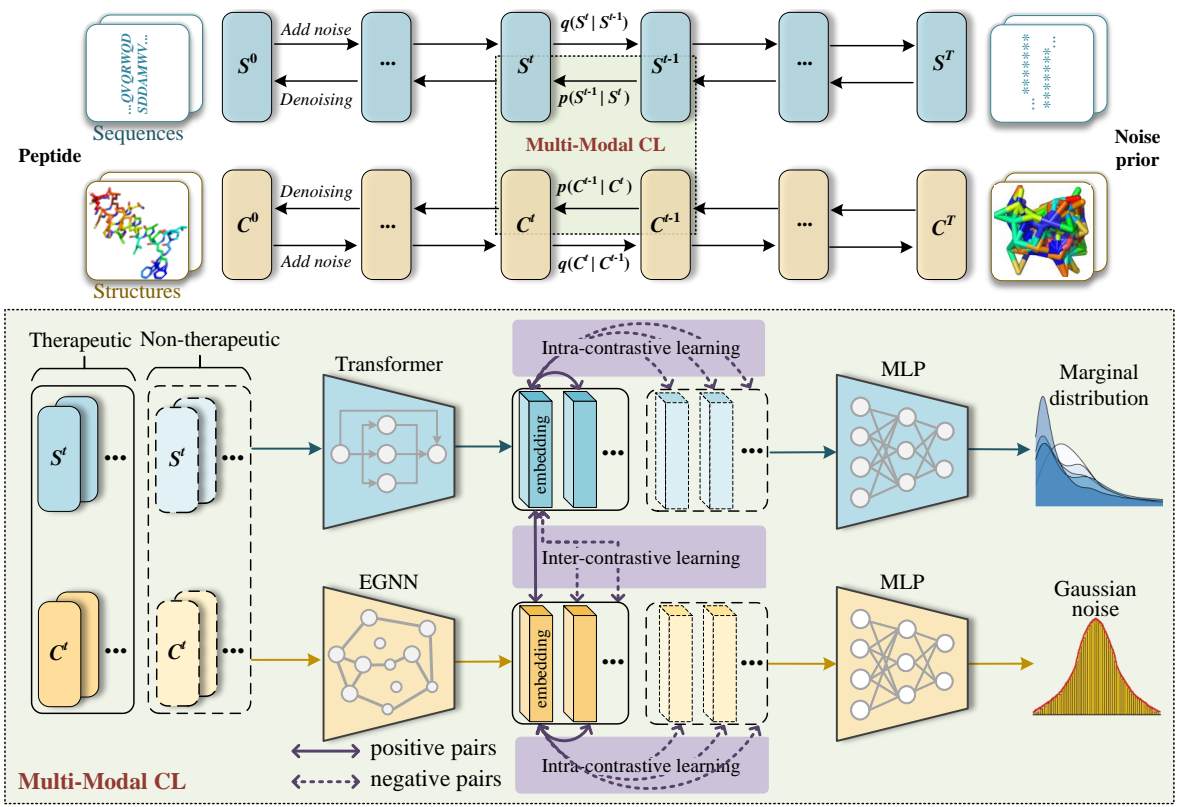}
\caption{Overview of the MMCD. MMCD consists of a diffusion model for the peptide sequence-structure co-generation and multi-modal contrastive learning (CL). The diffusion model involves a forward process ($q(\cdot|\cdot)$) for adding noise and a reverse process ($p(\cdot|\cdot)$) for denoising at each timestep $t$. The reverse process utilizes a transformer encoder (or EGNN) to extract embeddings from sequences $S$ (or structures $C$), and a sequence (or structure)-based MLP to map embeddings to the marginal distribution (or Gaussian) noise. The multi-modal CL includes an Inter-CL and an Intra-CL, which aims to align sequence and structure embeddings, and differentiate therapeutic and non-therapeutic peptide embeddings.}
\end{figure*}

\section{Related works}
\subsection{Diffusion Model for Protein Generation}
Diffusion models \cite{songGenerativeModeling2019, trippeDiffusionProbabilistic2023} devote to learning the noise that adequately destroys the source data and iteratively remove noise from the prior distribution to generate new samples, which have emerged as cutting-edge methods for numerous generation tasks, especially in proteins \cite{wuProteinStructure2022,caoSurveyGenerative2023}. For example, \citet{liuTextguidedProtein2023} proposed a textual conditionally guided diffusion model for sequence generation. \citet{hoogeboomEquivariantDiffusion2022} introduced ProtDiff with an E(3) equivariant graph neural network to learn a diverse distribution over backbone coordinates of structures. \citet{luoAntigenSpecificAntibody2022} considered both the position and orientation of antibody residues, achieving an equivariant diffusion model for sequence-structure co-generation. Despite their success, the fusion of both sequence and structure modalities in diffusion models has not been comprehensively investigated, and their potential for peptide generation remains unexplored. To fill this gap, we implement a peptide-oriented diffusion model capable of sequence-structure co-generation and multi-modal data fusion.

\subsection{Contrastive Learning}
Being popular in self-supervised learning, contrastive learning (CL) allows models to learn the knowledge behind data without explicit labels \cite{xia2022fast, zhuDiscreteContrastive2023}. It aims to bring an anchor (i.e., data sample) closer to a positive/similar instance and away from many negative/dissimilar instances, by optimizing their mutual information in the embedding space. Strategies to yield the positive and negative pairs often dominate the model performance \cite{zhang2022label}. For example, \citet{yuan2021multimodal} proposed a multi-modal CL to align text and image data, which encourages the agreement of corresponding text-image pairs (positive) to be greater than those of all non-corresponding pairs (negative). \citet{wu2022mitigating} designed a CL framework that makes full use of semantic relations among text samples via efficient positive and negative sampling strategies, to mitigate data sparsity for short text modeling. \citet{zhangPreTrainingProtein2023} augmented the protein structures using different conformers, and maximized the agreement/disagreement between the learned embeddings of same/different proteins, aiming to learn more discriminative representations. However, these CL strategies have yet to be extended to peptide-related studies. Therefore, we devise the novel CL strategy in peptide generation, which serves as an auxiliary objective to enforce sequence-structure alignment and boost model performance. 

\section{Methodology}
In this section, we formulate the peptide co-generation problem for sequence and structure. Subsequently, we elaborately enumerate the components of our method MMCD, including the diffusion model for peptide generation and the multi-modal contrastive learning strategy. The overview of MMCD is illustrated in Figure 1.

\subsection{Problem Formulation}
A peptide with $N$ residues (amino acids) can be represented as a sequence-structure tuple, denoted as $X=(S,C)$. $S=[s_i]_{i=1}^{N}$ stands for the sequence with $s_i\in \{ACDEFGHIKLMNPQRSTVWY\}$ as the type of the $i$-th residue, and $C=[c_i]_{i=1}^{N}$ stands for the structure with $c_i \in \mathbb{R}^{3\ast4}$ as Cartesian coordinates of the $i$-th residue (involving four backbone atoms N-C$_\alpha$-C-O). Our goal is to model the joint distribution of $X$ based on the known peptide data, so that sequences (i.e., residue types) and structures (i.e., residue coordinates) of new peptides can be co-generated by sampling the distribution.

\subsection{Diffusion Model for Peptide Generation}
The diffusion model defines the Markov chains of processes, in which latent variables are encoded by a \emph{forward diffusion process} and decoded by a \emph{reverse generative process} \cite{sohl-dicksteinDeepUnsupervised2015}. Let $X^0=(S^0,C^0)$ denotes the ground-truth peptide and $X^t=(S^t,C^t)$ for $t=1,...,T$ to be the latent variable at timestep $t$. The peptide generation can be modeled as an evolving thermodynamic system, where the forward process $q(X^t|X^{t-1})$ gradually injects small noise to the data $X^0$ until reaching a random noise distribution at timestep $T$, and the reverse process $p_\theta(X^{t-1}|X^t)$ with learnable parameters $\theta$ learns to denoise the latent variable $X^t$ towards the data distribution \cite{luoAntigenSpecificAntibody2022}.

\subsubsection{Diffusion for Peptide Sequence.}
Following \citet{anandProteinStructure2022}, we treat residue types as categorical data and apply discrete diffusion to sequences, where each residue type is characterized using one-hot encoding with 20 types. For the forward process, we add noise to residue types using the transition matrices with the marginal distribution \cite{austinStructuredDenoising2021, vignacDiGressDiscrete2023} (see details in Appendix A). For the reverse process, the diffusion trajectory is parameterized by the probability $q(S^{t-1}\mid S^t,S^0)$ and a network ${\hat{p}}_\theta$ is defined to predict the probability of $S^0$ \cite{austinStructuredDenoising2021}, that is:
\begin{equation}
p_\theta\big(S^{t-1}\mid S^t\big)=\prod_{1\leq i\leq N}q(s_i^{t-1}\mid S^t,\hat{S}^0)\cdot\hat{p}_\theta(\hat{S}^0\mid S^t)
\end{equation}
where $s_i^t$ denotes the one-hot feature for the $i$-th residue in the sequence $S$ at timestep $t$, and $\hat{S}^0$ is the predicted probability of $S^0$. In this work, we design the ${\hat{p}}_\theta$ as follows:
\begin{equation}
\hat{p}_{\theta}\left(\hat{S}^{0}\mid S^{t}\right)=\prod_{1\leq i\leq N}\text{Softmax}\left(\hat{s}_{i}^{0}\mid \mathcal{F}_s\left(h_{i}^{t}\right)\right)
\end{equation}
where $h_{i}^t$ is the input feature of residue $i$ with the diffusion noise at time $t$ (the initialization of $h_{i}^t$ is provided in Appendix A). $\mathcal{F}_s$ is a hybrid neural network to predict the noise of residue types from the marginal distribution, and then the noise would be removed to compute the probability of $\hat{s}_i^0$. Softmax is applied over all residue types. Here, we implement $\mathcal{F}_s$ with a transformer encoder and an MLP. The former learns contextual embeddings of residues from the sequence, while the latter maps these embeddings to the noises of residue types. The learned sequence embedding (defined as $\mathcal{S}$) involves downstream contrastive learning strategies.

\subsubsection{Diffusion for Peptide Structure.}
As the coordinates of atoms are continuous variables in the 3D space, the forward process can be defined by adding Gaussian noise to atom coordinates \cite{hoDenoisingDiffusion2020} (see details in Appendix A). Following \citet{trippeDiffusionProbabilistic2023}, the reverse process can be defined as: 
\begin{align}
p_\theta(c_i^{t-1}\mid C^t)&=\mathcal{N}(c_i^{t-1}\mid\mu_\theta(C^t,t),\beta^t{I})\\
\mu_\theta\left(C^t,t\right)&=\frac{1}{\sqrt{\alpha^t}}\left(c_i^t-\frac{\beta^t}{\sqrt{1-{\overline{\alpha}}^t}}\epsilon_\theta\left(C^t,t\right)\right)
\end{align}
where $c_i$ refers to coordinates of the $i$-th residue in the structure $C$; $\beta$ is the noise rate, formally $\alpha^t=1-\beta^t$, ${\overline{\alpha}}^t=\prod_{\tau=1}^{t}{(1-\beta^\tau)}$; the network $\epsilon_\theta$ is used to gradually recover the structural data by predicting the Gaussian noise. In this work, we design the $\epsilon_\theta$ as follows:
\begin{equation}
\epsilon_\theta(C^t,t)=\mathcal{F}_c\left(r_i^t,h_i^t\right)
\end{equation}
where $r_i$ represents the coordinates of residue $i$, $h_i$ is the residue feature, and $\mathcal{F}_c$ is a hybrid neural network for predicting Gaussian noises at timestep $t$. Similar to sequence diffusion, we implement $\mathcal{F}_c$ with an equivariant graph neural network (EGNN) \cite{satorrasEquivariantGraph2021} and an MLP. The former learns spatial embeddings of residues from the structure (formalized as a 3D graph), while the latter maps these embeddings to Gaussian noises. The learned structure embedding (defined as $\mathcal{C}$) also involves downstream contrastive learning strategies.

\subsubsection{Diffusion Objective.}
Following previous work \cite{anandProteinStructure2022}, we decompose the objective of the peptide diffusion process into sequence loss and structure loss. For the sequence loss $\mathcal{L}_S^t$, we aim to minimize the cross-entropy ($\text{CE}$) loss between the actual and predicted residue types at timestep $t$:
\begin{equation}
\mathcal{L}_S^t=\frac{1}{N}\sum_{1\leq i\leq N}\text{CE}\left(s_i^0,\hat{p}_\theta(\hat{s}_i^0|S^t)\right)
\end{equation}

For the structure loss $\mathcal{L}_C^t$, the objective is to calculate the mean squared error (MSE) between the predicted noise $\epsilon_\theta$ and standard Gaussian noise $\epsilon$ at timestep $t$:
\begin{equation}
\mathcal{L}_C^t=\frac{1}{N}\sum_{1\leq i\leq N}\left\|\epsilon_i-\epsilon_\theta(C^t,t)\right\|^2
\end{equation}

\subsection{Multi-Modal Contrastive Learning Strategy}
When multiple modal data (e.g., sequence and structure) coexist, it becomes imperative to capture their consistency to reduce the heterogeneous differences between modalities, allowing them to be better fused in generation tasks. Mutual information (MI) is a straightforward solution to measure the non-linear dependency (consistency) between variables \cite{liuTextguidedProtein2023}; thus, maximizing MI between modalities can force them to align and share more crucial information. Along this line, we bring in contrastive learning (CL) to align sequences and structures by maximizing their MI in the embedding space. Specifically, we devise CL strategies for each diffusion timestep $t$, as follows:

\subsubsection{Inter-CL.} For a peptide, we define its sequence as the anchor, its structure as the positive instance, and the structures of other peptides in a mini-batch as the negative instances. Then, we maximize the MI of positive pair (anchor and positive instance) while minimizing the MI of negative pairs (anchor and negative instances), based on embeddings learned from the networks ${\hat{p}}_\theta$ and $\epsilon_\theta$. Further, we establish a 'dual' contrast where the structure acts as an anchor and sequences are instances. The objective is to minimize the following InfoNCE-based \cite{chen2020simple} loss function:
{\small  {\begin{equation}
\mathcal{L}_\text{inter}^t=-\frac{1}{2}\left[\log\frac{E\left( \mathcal{S}_i^t,\mathcal{C}_i^t\right)}{\sum_{j=1}^ME\left(\mathcal{S}_i^t,\mathcal{C}_j^t\right)}
+\log\frac{E\left(\mathcal{C}_i^t,\mathcal{S}_i^t\right)}{\sum_{j=1}^{M}E\left(\mathcal{C}_i^t,\mathcal{S}_j^t\right)}\right]
\end{equation}}} 
where $\mathcal{S}_i/\mathcal{C}_i$ is the sequence/structure embeddings of $i$-th peptide in the mini-batch, $E(\cdot,\cdot)$ is the cosine similarity function with the temperature coefficient to measure the MI score between two variables, $M$ is the size of a mini-batch.

In addition, the used diffusion model can only remember confined generation patterns if therapeutic peptide data for training is limited, which may lead to inferior generalization towards novel peptides. To alleviate this issue, we introduce contrastive learning to boost the generative capacity of networks ${\hat{p}}_\theta$ and $\epsilon_\theta$ by enriching the supervised signals. However, it is unwise to construct positive instances by performing data augmentations on therapeutic peptides, as even minor perturbations may lead to significant functional changes \cite{yadav2022structural}. Hence, our focus lies on employing effective strategies for selecting negative instances. In this regard, we collect non-therapeutic peptides from public databases to treat them as negative instances, and maximize the disagreement between embeddings of therapeutic and non-therapeutic peptides. In detail, we devise an Intra-CL strategy for each diffusion timestep $t$, as follows:

\begin{table*}[ht]
\centering
\begin{threeparttable}[b]
\setlength{\tabcolsep}{10pt}
\begin{tabular}{ccccccc}
\hline
\multirow{2}{*}{Methods} & \multicolumn{3}{c}{AMP}                               & \multicolumn{3}{c}{ACP}                  \\ \cmidrule(r){2-4} \cmidrule(r){5-7}
                         & Similarity↓      & Instability↓     & Antimicrobial↑  & Similarity↓ & Instability↓ & Anticancer↑ \\ \hline
LSTM-RNN                 & 39.6164          & 45.0862          & 0.8550          & 36.9302     & 47.0669      & 0.7336      \\
$\text{AMPGAN}^*$        & 38.3080          & 51.5236          & 0.8617          & -           & -            & -           \\
$\text{HydrAMP}^*$       & 31.0662          & 59.6340          & 0.8145          & -           & -            & -           \\
$\text{WAE-PSO}^*$       & -                & -                & -               & 41.2524     & 42.5061      & 0.7443      \\
DiffAB                   & 28.9849          & 43.3607          & 0.8024          & 31.4220     & 36.0610      & 0.6669      \\
SimDiff                  & 25.5385          & 41.1629          & 0.8560          & 28.8245     & 33.0405      & 0.7222      \\ 
\textbf{MMCD}            & \textbf{24.4107} & \textbf{39.9649} & \textbf{0.8810} & \textbf{27.4685} & \textbf{31.7381} & \textbf{0.7604}  \\ \hline
\end{tabular}
\begin{tablenotes}
\item '*' represents that the method relies on domain-specific biological knowledge. '-' represents that the method is unsuitable for the current task. For example, AMPGAN and HydrAMP are only designed for the AMP generation.
\end{tablenotes}
\end{threeparttable}
\caption{Results for the sequence generation} 
\end{table*}

\begin{table*}[ht]
\centering
\setlength{\tabcolsep}{16pt}
\begin{tabular}{cccccc}
\hline
\multirow{2}{*}{Methods} & \multicolumn{3}{c}{AMP}                            & \multicolumn{2}{c}{ACP} 
\\ \cmidrule(r){2-4} \cmidrule(r){5-6}
                         & Ramachandran↑    & RMSD↓           & Docking↑        & Ramachandran↑  & RMSD↓  \\ \hline
APPTEST                  & 69.6576          & 2.7918          & 1362            & 67.9826        & 2.8055 \\
FoldingDiff              & 72.4681          & 2.5118          & 1574            & 72.0531        & 2.6033 \\
ProtDiff                 & 71.3078          & 2.5544          & 1533            & 69.7589        & 2.4960 \\
DiffAB                   & 72.9647          & 2.3844          & 1608            & 71.3225        & 2.5513 \\
SimDiff                  & 76.1378          & 2.1004          & 1682            & 76.6164        & 2.4118 \\ 
\textbf{MMCD}            & \textbf{80.4661} & \textbf{1.8278} & \textbf{1728}   & \textbf{78.2157}        & \textbf{2.0847} \\ \hline
\end{tabular}
\caption{Results for the structure generation.} 
\end{table*}

\subsubsection{Intra-CL.} 
In a mini-batch, we define the sequence of a therapeutic peptide $i$ as the anchor, and the sequence of another therapeutic peptide $j$ as the positive instance, while the sequences of non-therapeutic peptides $k$ are regarded as negative instances. Similar to Inter-CL, we then maximize/minimize the MI of positive/negative pairs. And we also establish a structure-oriented contrast by using structures of therapeutic and non-therapeutic peptides to construct the anchor, positive, and negative instances. The objective is to minimize the following loss function \cite{zhengWeaklySupervised2021}: 
\begin{equation}
\begin{aligned}
\mathcal{L}_\text{intra}^t&=-\frac{1}{M} \sum_{j=1, j\neq i}^{M}1_{y_i= y_j} 
\left(\log\frac{E\left( \mathcal{S}_i^t,\mathcal{S}_j^t\right)}{\sum_{k=1}^M 1_{y_i\neq y_k} E\left(\mathcal{S}_i^t,\mathcal{S}_k^t\right)}\right.\\
&\left.+\log\frac{E\left(\mathcal{C}_i^t,\mathcal{C}_j^t\right)}{\sum_{k=1}^{M}1_{y_i\neq y_k} E\left(\mathcal{C}_i^t,\mathcal{C}_k^t\right)}\right)
\end{aligned}
\end{equation}
where $y_i$ represents the class of peptide $i$ (i.e., therapeutic or non-therapeutic). $1_{y_i= y_j}$ and $1_{y_i\neq y_k}$ stand for the indicator functions, where the output is $1$ if $y_i=y_j$ (peptides $i$ and $j$ belong to the same class) or $y_i \neq y_k$ (the types of peptides $i$ and $k$ are different); otherwise the output is $0$. The indicator function filters therapeutic and non-therapeutic peptides from the data for creating positive and negative pairs.

The reason behind the design of Intra-CL is intuitive. First, the non-therapeutic class naturally implies opposite information against the therapeutic class, and hence it makes the model more discriminative. Second, the fashion to maximize the disagreement between classes (1) can induce biases in the embedding distribution of therapeutic peptides, identifying more potential generation space, and (2) can explicitly reinforce embedding-class correspondences during diffusion, maintaining high generation fidelity \cite{zhu2022discrete}. Further analysis is detailed in the ablation study.

\subsection{Model Training}
The ultimate objective function is the sum of the diffusion process for sequence and structure generation, along with the CL tasks for Intra-CL and Inter-CL:
{\small
\begin{equation}
\mathcal{L}_\text{total}=\mathbb{E}_{t\thicksim \text{Uniform(1...T)}}\left[\alpha\left(\mathcal{L}_S^t+\mathcal{L}_C^t\right)+\left(1-\alpha\right)\left(\mathcal{L}_\text{intra}^t+\mathcal{L}_\text{inter}^t\right)\right]
\end{equation}
}where $\alpha$ represents a hyperparameter to balance the contributions of different tasks. The $\text{Uniform(1...T)}$ shows the uniform distribution for the diffusion timesteps. The implementation details of MMCD and the sampling process of peptide generation can be found in Appendix A.

\section{Experiments}
\begin{figure*}[ht]
\centering
\includegraphics[scale=0.4]{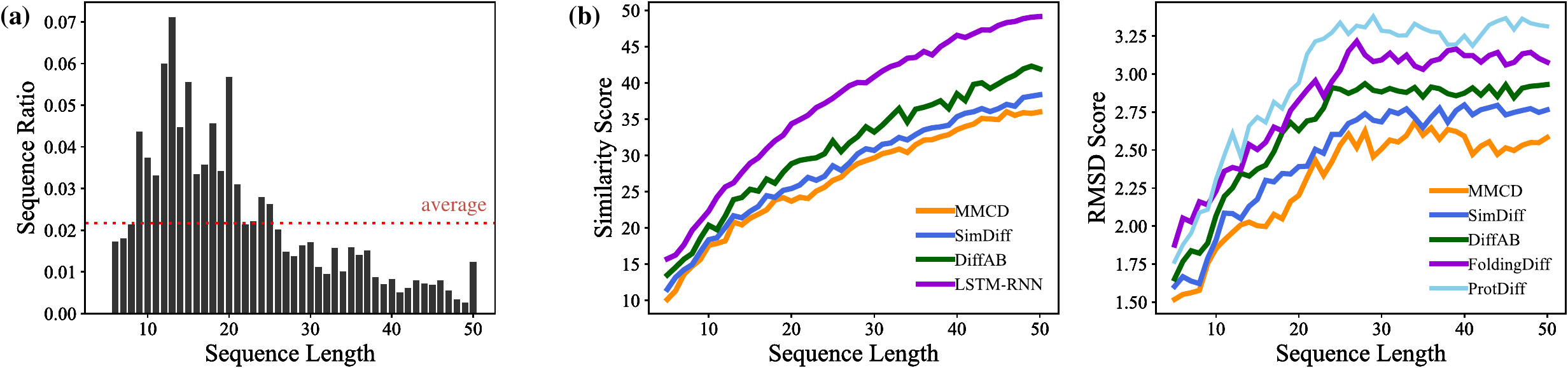}
\caption{(a) The sample ratio under different sequence lengths in the AMP dataset, where the red line is the average ratio. (b) The similarity and RMSD scores of MMCD and baselines across different sequence lengths.}
\end{figure*}

\subsection{Experimental Setups}
\subsubsection{Datasets.}
Following previous studies \cite{thiphanMLACPUpdated2022, zhangDeepLearningBased2023}, we collected therapeutic peptide data from public databases, containing two biological types, i.e., antimicrobial peptides (AMP) and anticancer peptides (ACP). Among these collected peptides, a portion of them only have 1D sequence information, without 3D structure information. Then, we applied Rosetta-based computational tools \cite{chaudhuryPyRosettaScriptbased2010} to predict the missing structures based on their sequences. Finally, we compiled two datasets, one containing 20,129 antimicrobial peptides and the other containing 4,381 anticancer peptides. In addition, we paired an equal number of labeled non-therapeutic peptides (collected from public databases) with each of the two datasets, exclusively for the contrastive learning task.

\subsubsection{Baselines.}
We compared our method with the following advanced methods for peptide generation at sequence and structure levels. For the sequence generation, the autoregression-based method LSTM-RNN \cite{mullerRecurrentNeural2018}, the GAN-based method AMPGAN \cite{oortAMPGANV2Guided2021}, and the VAE-based methods including WAE-PSO \cite{yangAcceleratingDiscovery2022} and HydrAMP \cite{szymczakDiscoveringHighly2023} are listed as baselines. For the structure generation, we took APPTEST \cite{timmonsAPPTESTNovel2021} as a baseline, which combines the neural network and simulated annealing algorithm for structure prediction. Moreover, we extended diffusion-based methods for protein generation to peptides. The diffusion-based methods for structure generation (e.g., FoldingDiff \cite{wuProteinStructure2022} and ProtDiff \cite{trippeDiffusionProbabilistic2023}) and the sequence-structure co-design (e.g., DiffAB\cite{luoAntigenSpecificAntibody2022} and SimDiff\cite{zhangPreTrainingProtein2023}), are considered for the comparison separately in the sequence and structure generation. 

\subsubsection{Evaluation protocol.}
Here, we required each model (ours and baselines) to generate 1,000 new peptides, and then evaluated the quality of generated peptides with the following metrics. For the sequence, \textbf{similarity} score is used to quantify how closely the generated sequences match existing ones, with a lower score indicating higher novelty; \textbf{instability} score \cite{mullerModlAMPPython2017} indicates the degree of peptide instability; \textbf{antimicrobial}/\textbf{anticancer} score evaluates the probability of peptides having therapeutic properties. For the structure, \textbf{Ramachandran} score \cite{hollingsworthFreshLook2010} accesses the reliability of peptide structures; \textbf{RMSD} score measures the structural similarity between generated and existing peptides, with a lower score indicating higher authenticity; \textbf{docking} score \cite{florez-castilloIbM6Antimicrobial2020} evaluates the binding degree of antimicrobial peptides to bacterial membrane proteins (PDB ID: 6MI7). We only reported the average metrics over all generated peptides for each method in the experimental results. Detailed information about the datasets, baselines, metrics, and implementations can be found in Appendix B. Our code, data and appendix are available on GitHub (https://github.com/wyky481l/MMCD)

\subsection{Experimental Results}
\subsubsection{Performance comparison.}
In the results of sequence generation under two datasets (as shown in Table 1), MMCD exhibited lower similarity and instability scores than all baselines, suggesting its good generalization ability in generating diverse and stable peptides. Meanwhile, MMCD surpassed all baselines with higher antimicrobial and anticancer scores across AMP and ACP datasets, highlighting its strong potential for generating therapeutic peptides. Beyond that, we noticed that diffusion-based baselines (e.g., SimDiff, DiffAB) exhibit higher stability and diversity but lower therapeutic scores compared to baselines that incorporate biological knowledge (e.g., AMPGAN, HydrAMP, WAE-PSO, details in Appendix B). By contrast, MMCD introduced biological knowledge into the diffusion model by designing the contrastive learning of therapeutic and non-therapeutic peptides, thereby delivering optimality across various metrics.

For the results of structure generation (as shown in Table 2), MMCD also outperformed all the baselines and exceeded the best baselines (DiffAB and SimDiff) by 23.3$\%$ and 12.9$\%$ in RMSD scores, 10.2$\%$ and 5.6$\%$ in Ramachandran scores, and 7.4$\%$ and 2.7$\%$ in docking scores for AMP dataset. The higher Ramachandran score and lower RMSD score of MMCD underlined the reliability of our generated peptide structures. Especially in peptide docking, we found that MMCD shows the best docking score compared with baselines, which indicates great binding interactions with the target protein. Overall, MMCD is superior to all baselines in both sequence and structure generation of peptides, and its impressive generative ability holds great promise to yield high-quality therapeutic peptides.

\begin{table*}[ht]
\centering
\setlength{\tabcolsep}{6pt}
\begin{tabular}{ccccccc}
\hline
\multirow{2}{*}{Methods} & \multicolumn{3}{c}{AMP}                     & \multicolumn{3}{c}{ACP}                  \\ \cmidrule(r){2-4} \cmidrule(r){5-7}
                         & Similarity↓ & Instability↓ & Antimicrobial↑ & Similarity↓ & Instability↓ & Anticancer↑ \\ \hline
MMCD (w/o InterCL \& IntraCL)  & 27.4794     & 42.5359      & 0.8013         & 31.2820     & 34.6888      & 0.6996      \\
MMCD (w/o IntraCL)             & 26.6889     & 41.2631      & 0.8584         & 28.9782     & 33.0268      & 0.7513      \\
MMCD (w/o InterCL)             & 24.9079     & 41.7646      & 0.8494         & 28.0143     & 33.9816      & 0.7352      \\
MMCD                           & 24.4107     & 39.9649      & 0.8810         & 27.4685     & 31.7381      & 0.7604      \\ \hline
\end{tabular}
\caption{Ablation study on the sequence-level generation task.} 
\end{table*}

\subsubsection{Performance on different sequence lengths.}
In our dataset, sequence lengths of different peptides exhibited substantial variation, with the number of residues ranging from 5 to 50 (Figure 2-a). We required models to generate 20 new peptides (sequences or structures) at each sequence length. Note that two methods, AMPGAN and HydrAMP, were excluded from the comparison because they cannot generate peptides with fixed lengths. From the generated results on the AMP dataset (Figure 2-b), MMCD exceeded the baselines in terms of similarity and RMSD scores at each sequence length. With the increasing sequence lengths, there is a general trend of increased similarity and RMSD scores across all methods. One possible reason for this trend is that designing longer peptides becomes more complex, given the more prominent search space involved. Additionally, the scarcity of long-length peptides poses challenges in accurately estimating the similarity between generated and known peptides. In summary, these observations supported that MMCD excels at generating diverse peptides across different lengths, especially shorter ones.

\begin{figure}[h]
\includegraphics[scale=0.3]{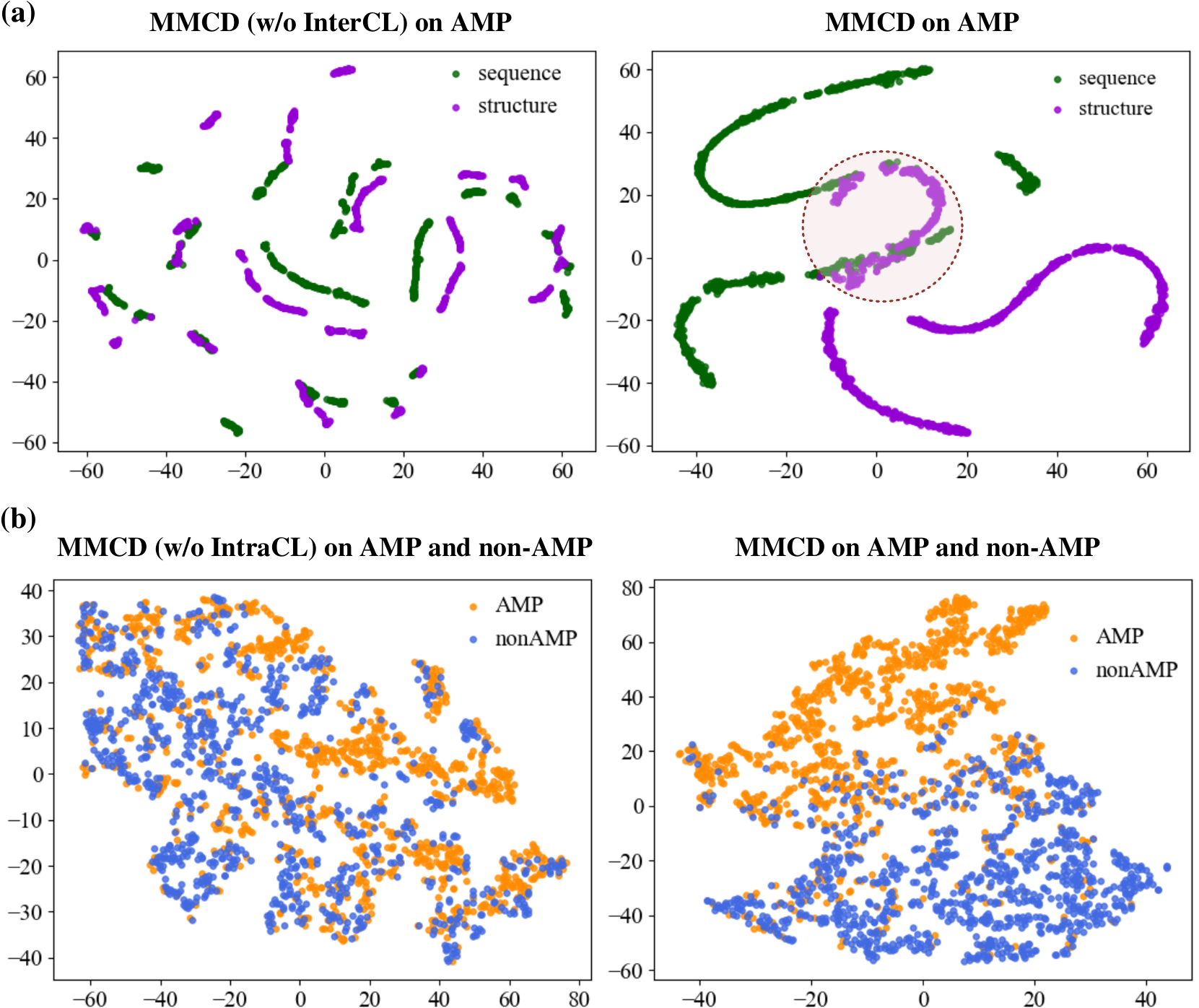}
\caption{(a) The t-SNE for structure and sequence embeddings of therapeutic peptides (AMP data) obtained from MMCD (w/o Inter-CL) and MMCD. (b) The t-SNE for embeddings (including structures and sequences) of therapeutic (AMP) and non-therapeutic (non-AMP) peptides obtained from MMCD (w/o Intra-CL) and MMCD.}
\end{figure}

\subsection{Ablation study}
To investigate the necessity of each module in MMCD, we conducted several comparisons between MMCD with its variants: (1) MMCD (w/o Inter-CL) that removes the Inter-CL task, (2) MMCD (w/o Intra-CL) that removes the Intra-CL task, and (3) MMCD (w/o Inter-CL \& Intra-CL) that removes both Inter-CL and Intra-CL tasks. The comparisons were operated on both AMP and ACP datasets, and the results are shown in Table 3 and Appendix Table 1. When the Inter-CL was removed (w/o Inter-CL), we observed a decline in all metrics for peptide sequence and structure generation, implying the importance of aligning two modalities via CL. The variant (w/o Intra-CL) results signified that using the CL to differentiate therapeutic and non-therapeutic peptides contributes to the generation. As expected, the performance of MMCD dropped significantly after removing both Inter-CL and Intra-CL (w/o Inter-CL \& Intra-CL). 

To better understand the strengths of Inter-CL and Intra-CL, we performed the t-SNE \citep{tsne} visualization using the learned embeddings of peptides on the AMP dataset. As illustrated in Figure 3-a, Inter-CL effectively promoted the alignment of sequence and structure embeddings, facilitating the shared crucial information (dashed circle) to be captured during diffusion. The t-SNE of Intra-CL (Figure 3-b) also revealed that it better distinguished therapeutic peptides from non-therapeutic ones in the embedding distribution. And the resulting distribution bias may identify more potential generation space, thus leading to higher quality and diversity of therapeutic peptides generated by MMCD. Overall, MMCD with all the modules fulfilled superior performance, and removing any modules will diminish its generation power. 

\subsection{Peptide-docking analysis}
To test the validity of generated peptide structures, we conducted a molecular-docking simulation. Here, a peptide was randomly selected from the AMP dataset as the reference, and the methods (Figure 4) were employed to generate corresponding structures based on the sequence of the reference peptide (see details in Appendix C). The lipopolysaccharide on the outer membrane of bacteria \cite{liStructuralBasis2019} was selected as the target protein for molecular docking. Then, we extracted the residues within a 5Å proximity between peptides (i.e., the reference and generated structures) and the active pocket of target protein in docking complexes, to visualize their binding interactions \cite{millerReliableAccurate2021}. Of these docking results, all methods yielded a new structure capable of binding to the target protein, and our method exhibited the highest docking scores and displayed binding residues most similar to the reference structure. This prominent result underscored the reliability and therapeutic potential of our method for peptide generation.

\begin{figure}[ht]
\includegraphics[scale=0.135]{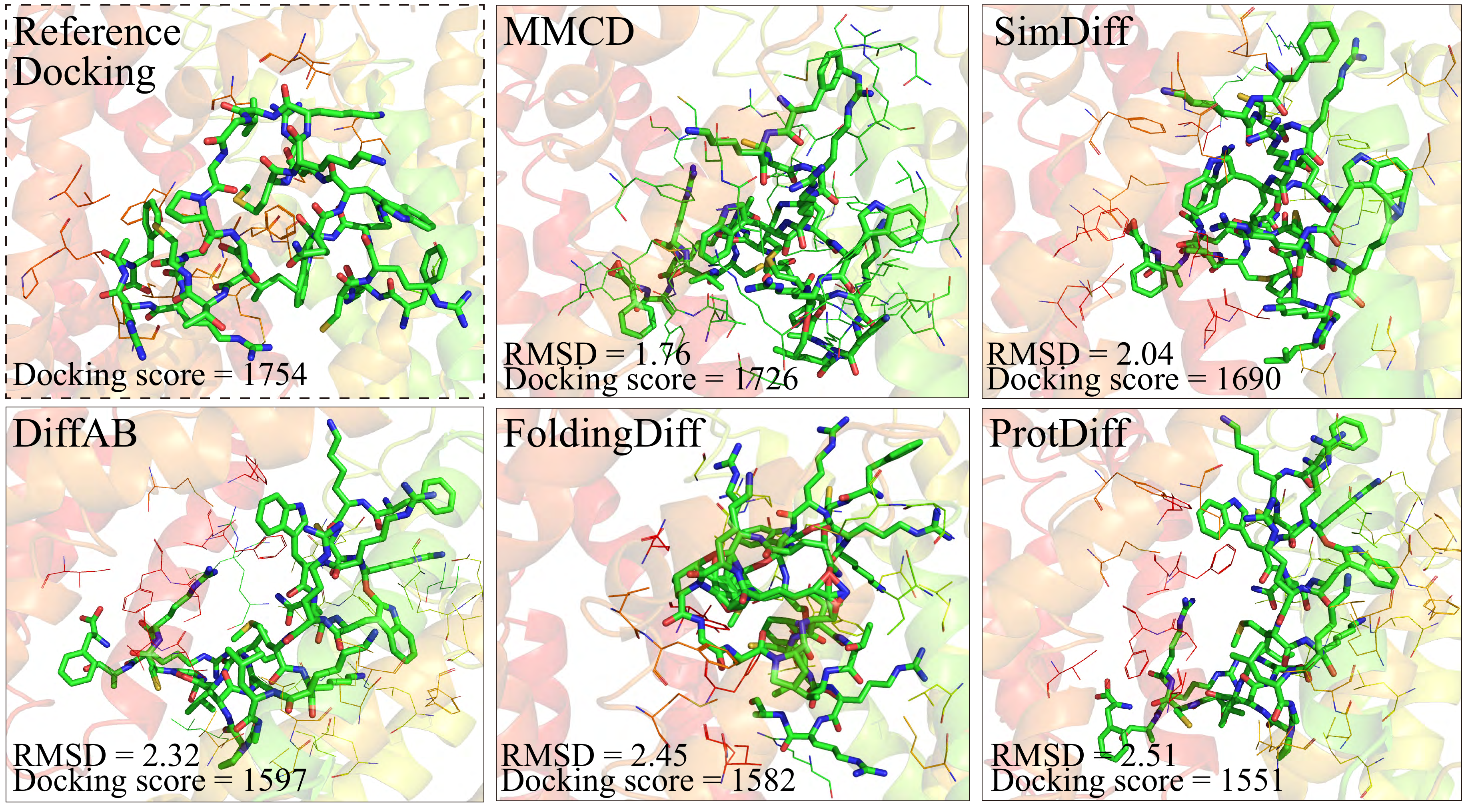}
\caption{Docking analysis (interactive visualization between target protein and peptides) of the reference and generated structures by MMCD and baselines. Thick lines represent the residues of peptides, and the thin lines show the binding residues for protein-peptide complexes.}
\end{figure}

\subsection{Conclusion}
In this work, we propose a multi-modal contrastive diffusion model for the co-generation of peptide sequences and structures, named MMCD. MMCD is dedicated to leveraging a multi-modal contrastive learning strategy to capture consensus-related and difference-related information behind the sequences/structures and therapeutic/non-therapeutic peptides, enhancing the diffusion model to generate high-quality therapeutic peptides. The experimental results unequivocally demonstrate the capability of our method in co-generating peptide sequence and structure, surpassing state-of-the-art baseline methods with advantageous performance.

\subsection{Acknowledgments}
This work was supported by the National Natural Science Foundation of China (62372204, 62072206, 61772381, 62102158); Huazhong Agricultural University Scientific \& Technological Self-innovation Foundation; Fundamental Research Funds for the Central Universities (2662021JC008, 2662022JC004). The funders have no role in study design, data collection, data analysis, data interpretation, or writing of the manuscript.

\bibliography{manuscript}

\end{document}